+

# Resonance method of electric-dipole-moment measurements in storage rings

Yuri F. Orlov [1,2], William M. Morse[1], Yannis K. Semertzidis[1]

[1]*Brookhaven National Laboratory, Upton, NY 11973;* [2]*Laboratory for Elementary-Particle Physics, Cornell University, Ithaca NY 14853.*

**Abstract**

A "resonance method" of measuring the electric dipole moment (EDM) of nuclei in storage rings is described, based on two new ideas: (1) Oscillating particles' velocities in resonance with spin precession, and (2) alternately producing two sub-beams with different betatron tunes—one sub-beam to amplify and thus make it easier to correct ring imperfections that produce false signals imitating EDM signals, and the other to make the EDM measurement. PACS numbers: 29.20Dh, 21.20.Ky.

1. <u>Introduction</u>. Discovery of the intrinsic electric dipole moment (EDM) would indicate direct violation of T (and P)-symmetry. This letter proposes to measure nuclei EDM's in storage rings with an accuracy such that the observed EDM values will be bigger than Standard Model (SM) predictions but well in the frame of non-SM theories, SUSY the leading candidate among them. The experiment will therefore, in effect, be a test of the SM. The new method proposed in this letter belongs to the developing area of using storage rings especially designed to measure EDM's [1,2,3]. For lack of space we cannot review the publications on non-storage- ring EDM measurements, for example [4,5,6]. We only note that our new method differs in almost all respects from that described in [1]. Its advantages include a much smaller ring (because we can use much bigger magnetic fields) and greater ease of canceling false signals imitating EDM (because it is easier to observe oscillating perturbations than those constant in time).

The method is based on: (1) using forced oscillations of particles' velocities in resonance with the spin precession in order to expose EDM (hence the name "resonance method"), and (2) alternately producing two sub-beams with different betatron tunes (by alternating a lens gradient) such that false EDM signals resulting from the ring imperfections are amplified in one of the sub-beams by a factor $K \gg 1$, permitting the imperfections to be corrected without affecting the EDM in the other sub-beam (which is used for the EDM measurement). After the imperfections are corrected, the false signals carried by the sub-beam to be used for the EDM measurement fall below the designed statistical error for EDM.

Our simulations show that a slightly modified resonance method can be used for EDM measurement of nuclei having $|a| \gtrsim 1$, $a = (g-2)/2$, like the proton or Helium-3. For reasons of space we consider here only the deuteron EDM measurement ($a = -0.142988$) typical for nuclei with a small anomalous gyromagnetic ratio.

2. <u>Using resonance between velocity and anomalous magnetic moment</u>. Let a charged polarized particle rotate in a storage ring (Fig. 1) with revolution frequency $\omega_c = 2\pi v/L$, where v is velocity, and $L$ is orbit length. The frequency of the spin planar precession around this orbit is $\omega_a = (e/mc)aB = a\gamma\omega_c$, $\gamma = \varepsilon/mc^2$. In its restframe, the particle interacts with vertical magnetic field $\vec{B}' = \gamma\vec{B}$ and radial electric field $\vec{E}' = \gamma[\vec{v} \times \vec{B}]$; the interaction Hamiltonian $H = -(\vec{\mu}\vec{B}' + \vec{d}\vec{E}')$, where $\vec{d} \equiv (e\hbar/2mc)\eta \cdot \vec{s}$ is the particle electric dipole moment. If we now oscillate electric field $\vec{E}'$ in resonance with the spin planar precession, we will observe a slow buildup of the vertical polarization, $P_y \equiv \langle s_y \rangle$, proportional to $d$. (We use $x$ for radial, $y$ for vertical, and $z$ for longitudinal coordinates.) As shown below, the only way to produce a spin resonance torque proportional to $d$ with the help of such an electric field is to oscillate the particle labframe velocity $\vec{v}$, and not the magnetic field $\vec{B}$.

Consider the ring without perturbations (Fig. 1). As defined in [7], the equation for the polarization vector $\vec{P}$ (or spin operator $\vec{s}$) in the presence of the EDM (marked by subscript $e$), and valid for every spin, can be written as

$$d\vec{s}/dt = [\vec{s} \times (\vec{\omega}_a + \vec{\omega}_e)]; \quad \vec{\omega}_a = (e/mc)a\vec{B}, \quad \vec{\omega}_e = (e/2mc)\eta\left[\vec{E} - \vec{\beta}(\vec{\beta}\vec{E})\gamma/(\gamma+1) + (\vec{\beta} \times \vec{B})\right] \quad (1)$$

$\beta = v/c$. (For the spin-1/2 case, the term $-\vec{\beta}(\vec{\beta}\vec{E})\gamma/(\gamma+1)$ in $\vec{\omega}_e$ was recently derived from extended Dirac equations [8].) Ideally, we have no labframe radial electric field. We have a longitudinal electric field, since we use synchrotron stability. But this field is small compared with $[\vec{v} \times \vec{B}]$, so we can neglect it in the $\omega_e$. We can also neglect the EDM term in the equations for $s_x$ and $s_z$. Now, our $\vec{v}$ (and possibly $\vec{B}$) depends on time. By design, if the longitudinal spin component oscillates as $s_z = s_{z0}\cos(\omega_a t + \varphi_a)$, then every particle velocity in our ring will contain three (instead of the usual two) terms:

$$v = v_0 + (\Delta v)_{sy}\cos(\omega_{sy} t + \alpha_{sy}) + (\Delta v)_m \cos(\omega_m t + \varphi_m), \quad \omega_m \approx \omega_a, \quad \varphi_m \approx \varphi_a, \quad (2)$$

where subscript *sy* refers to the free synchrotron oscillations and *m* ("modulated") refers to the forced (therefore, coherent) synchrotron oscillations designed especially for the EDM resonance. From (1),

$$ds_y/dt = (e/2mc)\eta[\vec{s} \times [\vec{v}/c \times \vec{B}]]_y = -(e/2mc^2)\eta v_z(t)B_y(z(t),t)s_z(t). \quad (3)$$

We can neglect the $\eta^2$-terms in (3). Then, introducing the planar spin phase $\Phi = \left((ae/mc)\int_0^t B_y(z(t),t)dt + \varphi_a\right)$, we get:

$$\Delta s_y(t) = \int_0^t dt(ds_y/dt) = \eta\frac{s_z(0)}{2a}\int_0^t \sin\Phi\frac{dv}{dt}d\Phi + \text{oscillating terms.} \quad (4)$$

We see that magnetic field oscillations cannot affect the buildup of $s_y$ if $dv/dt$ does not oscillate in resonance with $\sin\Phi$. In the case of precise resonance $\omega_m = \omega_a$, $\varphi_m = \varphi_a$,

$$\Delta s_y(t) = \eta\frac{s_{z0}}{4a}(\Delta v_m/c)\omega_a t = \eta s_{z0}\gamma(\Delta v_m/c)\omega_C t/4. \quad (5)$$

The bigger $(\Delta v)_m$, the smaller statistical error $\sigma_d \equiv \sqrt{\langle(\delta d)^2\rangle}$,

$$\sigma_d(e\cdot cm/year) \approx 1.6 \times 10^3 \hbar(eV\cdot s)/PA\Delta v_m(m/s)\langle B(T)\rangle\sqrt{fN_{fill}T_{exp}(s)\tau_{coh}(s)}. \quad (6)$$

The following parameters seem to be achievable: Acceptance of $N_{fill} = 2 \times 10^{11}$ deuteron/fill having momentum $p_0 = 1.5 GeV/c$, as our preliminary analysis of the collective beam effects shows; initial horizontal polarization $P = 0.95$; superconducting $B_0 = 3T$, $\langle B \rangle = 1.8T$, $R_0 = 1.7m$; the slow extraction of deuterons onto an external polarimeter with $fA^2 = 0.01$, where $f$ is the polarimeter efficiency and $A$ the left-right scattering asymmetry caused by the nonzero $s_y$; spin coherence time $\tau_{coh} = 1000s$ (the prolongation of which is based mostly on using multipole lenses [9], which are not shown in Fig. 1); and the amplitude of the forced velocity modulation $(\Delta v)_m = 3.5 \times 10^6 m/s$. With these parameters we get $\sigma_d = 2.5 \times 10^{-29} e \cdot cm/year$ (that is, after $T_{exp} = 1.8 \times 10^7 s$ of the EDM measurement).

To achieve as big a $(\Delta v)_m$ as possible we need superconducting RF cavities, see Fig. 1, with total voltage $V_0 = 10\text{-}20$ MV/turn. Let this voltage oscillate as $\sin \omega_{RF} t$, with $\omega_{RF} = h \omega_C$, where $h \sim 20 - 40$. The eigenfrequency $\omega_{sy}$ of the free synchrotron oscillations must be chosen close to the g-2 frequency, $\omega_a$. For modulations of the forced oscillations $(\Delta v)_m \cos(\omega_a t + \varphi_a)$, we can add one more RF cavity oscillating as $V_1 \sin(\omega_1 t + \varphi_a)$, with $V_1 \ll V_0$, and $(\omega_1 - \omega_{RF}) = \omega_a \approx \omega_{sy}$. Then the beatings between $\omega_1$ and $\omega_{RF}$, supported by voltage $V_1$, will create the needed coherent part of the synchrotron oscillations. Simulations (and theoretical analyses) show that there exist two possible regimes with big coherent oscillations: a strongly linear regime (using, for example, a specially designed RF cavity [10] for the linearization of oscillations), or a strongly nonlinear regime with well-stabilized coherent oscillations [11].

3. The ring and its operations. We have considered several versions of resonance EDM rings and different optics. Fig. 1 represents the simplest version, in which we have only two semi-circular magnets, with an optional field gradient $\partial B / \partial R \neq 0$ and two straight sections, the left (LSS) and the right (RSS) sides in Fig. 1. At the LSS side, the dispersion of the closed orbits corresponding to different particle momenta, $\Delta x = D \Delta p / p$, is not zero, and there are only two lenses, $F$. At the RSS side, where most quadrupole lenses and all RF cavities are placed, $D = 0$. The ring is not symmetric: lenses denoted by $F$, $F_1$, $Q$, and $-F_2$ (a defocusing lens) are not equal to one another. $Q$, a special quadrupole at the RSS side whose gradient alternates in time, will be discussed later. Fig. 1 shows two different asymmetric closed orbits corresponding to two different particle momenta. Particles' betatron oscillations are performed around such closed orbits. The closed orbit in the RSS is the same for all momenta. This property cancels most of the undesirable spin-resonance effects caused by imperfections of the lattice elements placed along the RSS. The reason is that, in our method, only field perturbations oscillating with the (g-2) frequency $\omega_a$ can imitate the EDM. Therefore, if the closed orbit does not oscillate together with the forced velocity oscillations (whose frequency is $\omega_a$), then the undesirable rotations of lenses do not produce such perturbations. Their vertical shifts still do, but only if the lenses' chromaticity is not compensated. As for the RF's, analysis and simulations show that only inclinations of cavities in the vertical plane can produce a false EDM signal—which will be canceled, together with other effects, by the procedures described in the next section.

The asymmetry of the closed orbits has been confirmed by various beam-tracking simulations. If particles with different momenta initially move along the center line of the RSS, they have different radial coordinates and different angles after passing one of the semi-circular magnets. The angles are corrected by lenses $F$, after which the closed orbits corresponding to the different momenta become parallel to one another and to the orbit at the RSS. Obviously, the orbits converge at the RSS after finishing the turn. But if a particle having $p \neq p_0$ at the LSS initially moves along any line that does not correspond to this momentum, then its trajectory will not be closed after one turn since the lens setup is not symmetric.

With the given $F$ and $\partial B / \partial R$ of the magnets, the length of the straight sections is chosen such that momentum compaction factor $\alpha_p \equiv (\Delta L / L)/(\Delta p / p) = 1$. Here $\Delta L = L(p) - L(p_0)$, $L(p)$ is the length of the closed orbit for momentum $p$. If $n \equiv -R(\partial B / \partial R)/B = 1$ (our choice for the magnets), then the length of a straight section is $\sim 2R$. (It must be zero if $n = 0$.) If $\alpha_p = 1$, then $p/L = p_0/L_0$. Therefore, $a\gamma(p)\omega_C(p) = a\gamma(p_0)\omega_C(p_0)$, that is, $\omega_a(p) = \omega_a(p_0)$. This means that the g-2 frequency does not oscillate together with velocity oscillations, which eliminates the possibility of false resonances imitating the EDM at beam frequencies close to $k\omega_a$, $k>1$. (See [13] for another application of $\alpha_p = 1$.)

Gradient $\partial B/\partial x$ of lens $Q$ oscillates, $\langle \partial B/\partial x \rangle = g\cos N\omega_c t$, where $g$ is the gradient averaged over a particle's longitudinal and betatron oscillations and along lens $Q$. In the case of $N = h/2$ (where $h = \omega_{RF}/\omega_C =$ the number of bunches rotating in the ring), every two sequential bunches passing $Q$ get on the average exactly opposite gradient kicks. As a result, we have two sub-beams with different betatron frequencies: the usual ones corresponding to the best beam stability (the "normal" sub-beam) and ones very close to spin resonance $(\omega_y)_{res} = k\omega_c \pm \omega_a$ (the "sensitive" sub-beam). Here $\omega_y$ is the frequency of the vertical betatron oscillations. The lattice imperfections met by the particles during their rotations in the ring are practically the same for both sub-beams. At the same time, the orbit perturbation and a false EDM signal caused by these imperfections are greatly amplified in the sensitive sub-beam. This amplification will permit us to see and cancel these perturbations,. The external polarimeter precisely identifies individual bunches; thenthe bunches of normal sub-beams are used to measure the EDM, and the bunches of sensitive sub-beams are used to correct imperfections, as explained in the next sections. The needed $g$-value is around 20 gauss/cm if the length of the $Q\sim$ 0.5m. These parameters can be rather easily achieved by using a specially designed superconducting *RF* cavity [10].

The assumed equality of lattice imperfections for both sub-beams is correct because, at the level of the accuracy of our corrections (see the next section), the orbits' perturbations responsible for the false spin resonance, though very different for different sub-beams, are too small to change the lattice imperfections for any sub-beam.

4. <u>Amplification and correction of systematic errors (the "two sub-beams" procedure)</u>. We propose to correct systematic errors in three stages: (a) The usual accelerator physics methods of spin perturbation corrections. (b) Corrections using observations of obviously non-EDM $s_y$-buildups. Note, for example, that the $(ds_y/dt)$ corresponding to $d \gg 10^{-26} e\cdot cm$ in the deuteron case can be observed even during a single fill time of ~1000s. So if we already know that $d < 10^{-26} e \cdot cm$ and nevertheless see a $(ds_y/dt)$ corresponding to $d >> 10^{-26}$, we can introduce some counter-perturbations to halt this obviously non-EDM growth. All perturbations, including our counter-perturbations, are then corrected at the next, final stage of accuracy. (c) The "two sub-beams" procedure, probably to be combined with the clockwise↔counter-clockwise technique imitating the $t \leftrightarrow -t$ transformation explained in [1]. As already noted, the main systematic errors come from perturbative fields, $B_x$, $B_z$ and $E_y$ only if such perturbations are coupled with the $\Delta v/v$ oscillations designed for the EDM spin resonance. Because the ring is not homogeneous along the azimuth, the spin resonance frequencies are $\omega_a + k\omega_C$, with integer $k$. Most such perturbations can be expressed in terms of particle deviations $y=y(t)$ oscillating with these frequencies [12]. The magnitude of a false EDM signal is proportional to the amplitude of these $y$-oscillations, whose sensitivity to the imperfections depends on the closeness of their frequency, $\omega_y$, to $(\omega_y)_{res}$, $y_{res} \propto 1/[\omega_y^2 - (\omega_a + k\omega_C)^2]$. Due to its closeness to resonance, a sensitive sub-beam will carry a false EDM signal $K$-times larger than that carried by a normal sub-beam, $K >> 1$ being the ratio of the amplitudes of the sub-beams' vertical >betatron oscillations. These amplitudes are proportional to spin >resonance perturbations.

Let us assume that all imperfections are amplified more or less equally, which occurs when the vertical betatron oscillations are sufficiently smooth. The correction procedure in such a case is very simple. When we observe an amplified false EDM signal from the sensitive sub-beam, we must correct the (invisible but obviously existing) imperfections by using some counter-imperfections in such a way that, at some time $t$, the integral buildup of the spin signal carried by the sensitive sub-beam equals zero:

$$\int_0^t \left(\frac{ds_y}{dt}\right)_{edm} dt + K\int_0^t \left[\left(\frac{ds_y}{dt}\right)_{imp} - \left(\frac{ds_y}{dt}\right)_{cor}\right] dt = 0, \qquad (7)$$

where $imp \equiv$ imperfection, $cor \equiv$ correction. In fact, condition (7) defines time $t$. The spin signal carried by the normal sub-beam at this time is

$$\int_0^t \left(\frac{ds_y}{dt}\right)_{edm} dt + \int_0^t \left[\left(\frac{ds_y}{dt}\right)_{imp} - \left(\frac{ds_y}{dt}\right)_{cor}\right] dt = \Delta s_y(t). \qquad (8)$$

From (7) and (8),

$$(\Delta s_y)_{edm} \equiv \int_0^t \left(\frac{\Delta s_y}{dt}\right)_{edm} dt = \frac{K}{K-1} \Delta s_y(t) \approx \Delta s_y(t), \; K \gg 1. \qquad (9)$$

Note that this rather miraculous result, $(\Delta s_y)_{edm} \approx \Delta s_y$, does not depend on $K$'s actual value or sign. Moreover, (9) is valid for both a single time interval with property (7) and any sum of such intervals.

When the vertical betatron oscillations are not smooth, then different imperfections can have different $K$'s known a priori; we omit here analysis of such cases.

5. <u>Tensor interactions</u>. Since the deuteron has spin one, its Hamiltonian possesses quadrupole interactions and magnetic and electric tensor polarizabilities that are quadratic in spin. Correspondingly, equation (3) also acquires quadratic terms. However, it can be shown that after these terms are averaged over an arbitrary quantum state, they oscillate with frequencies $n\omega_a, \; n = 1,2,3,4$, and therefore are averaged in time. The terms not averaged over time and over beam parameter distributions fall below the accuracy $10^{-30} e \cdot cm$.

6. <u>Conclusion</u>. By using the resonance method of EDM measurement we can move far beyond the current experimental limits on the EDM's of nuclei ($< 3 \times 10^{-26} e \cdot cm$ for neutrons [6]) and enter the area of non-Standard Model predictions.

For useful discussions and some parallel explorations, we are grateful to: M. Bai, V.G. Baryshevsky, M. Blaskiewicz, A. Facco, F.J.M. Farley, G. Hoffstaetter, H. Huang, K.P. Jungmann, B. Khazin, I.B. Khriplovich, D. Lazarus, A. Lehrach, A. Luccio, W.W. MacKay, J.P. Miller, C.J.G. Onderwater, S. Orlov, V. Ptitsyn, S. Redin, B.L. Roberts, T. Roser, Y. Shatunov, A. Silenko, A. Sidorin, E. Stephenson, G. Venanzoni, and M. Zobov.

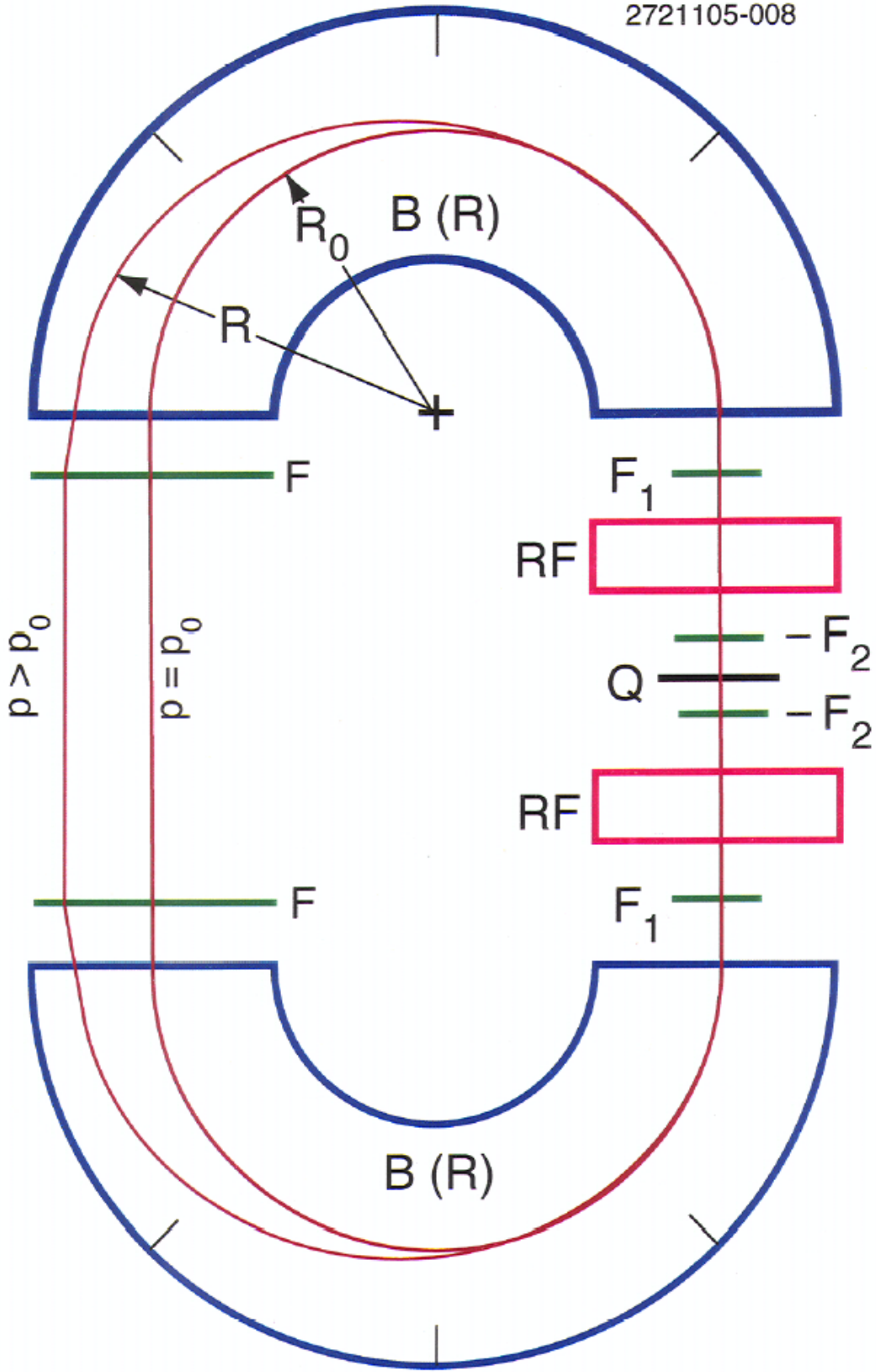

Figure 1. An EDM ring.